\documentclass[preprint,review,12pt,a4paper]{elsarticle}
\usepackage{graphicx}
\usepackage{epsf}
\usepackage{amsmath}
\usepackage{lscape}
\usepackage{color}
\usepackage{dcolumn}
\usepackage{rotating}
\usepackage{bm}
\usepackage{longtable}
\usepackage{color}
\biboptions{numbers,sort&compress}

\newcommand{\bra}[1]{\langle #1|}
\newcommand{\ket}[1]{|#1\rangle}

\newcommand{\p}{^\prime}

\journal{JQSRT}

\begin{document}
\begin{frontmatter}

\title{A global \textit{ab initio} dipole moment surface for methyl chloride}

\author[a,b]{Alec Owens}
\ead{alec.owens.13@ucl.ac.uk}
\author[a]{Sergei N. Yurchenko}
\author[a]{Andrey Yachmenev}
\author[a]{Jonathan Tennyson}
\author[b]{Walter Thiel}
\address[a]{Department of Physics and Astronomy, University College London, Gower Street, WC1E 6BT London, United Kingdom}
\address[b]{Max-Planck-Institut f\"{u}r Kohlenforschung, Kaiser-Wilhelm-Platz 1, 45470 M\"{u}lheim an der Ruhr, Germany}

\date{\today}

\begin{abstract}
A new dipole moment surface (DMS) for methyl chloride has been generated at the CCSD(T)/aug-cc-pVQZ(+d for Cl) level of theory. To represent the DMS, a symmetry-adapted analytic representation in terms of nine vibrational coordinates has been developed and implemented. Variational calculations of the infrared spectrum of CH$_3$Cl show good agreement with a range of experimental results. This includes vibrational transition moments, absolute line intensities of the $\nu_1$, $\nu_4$, $\nu_5$ and $3\nu_6$ bands, and a rotation-vibration line list for both CH$_{3}{}^{35}$Cl and CH$_{3}{}^{37}$Cl including states up to $J=85$ and vibrational band origins up to $4400{\,}$cm$^{-1}$. Across the spectrum band shape and structure are well reproduced and computed absolute line intensities are comparable with highly accurate experimental measurements for certain fundamental bands. We thus recommend the DMS for future use.
\end{abstract}


\end{frontmatter}

\section{Introduction}
\label{sec:intro}

 The proposal of methyl chloride as a potential biosignature gas~\citep{05SeKaMe.CH3Cl,13SeBaHu.CH3Cl,13aSeBaHu.CH3Cl} in the search for life outside of the Solar System has ignited interest in its infrared spectrum. There is now considerable motivation for a comprehensive rotation-vibration line list of CH$_3$Cl. Since methyl chloride is known to contribute to ozone depletion, any such line list would undoubtedly be useful in terrestrial studies. Its importance as an atmospheric molecule is confirmed by the huge number of recent spectroscopic studies~\citep{11BrPeJa.CH3Cl,12BrTrJa.CH3Cl,12BrJaBu.CH3Cl,13BrJaLa.CH3Cl,13aBrJaLa.CH3Cl,12GuRoBu.CH3Cl,11BuGuEl.CH3Cl,12BuRoxx.CH3Cl,13Buxxxx.CH3Cl,13BuMaMo.CH3Cl,13RaJaDh.CH3Cl,14RaJaDh.CH3Cl,14aRaJaDh.CH3Cl,13DuLaBu.CH3Cl,14DuLaBu.CH3Cl,03NiFeCh.CH3Cl,04NiChBu.CH3Cl,05NiChxx.CH3Cl,05NiChBu.CH3Cl,16NiDmGo.CH3Cl,15GuRoMa.CH3Cl,15SaZhAl.CH3Cl,16LuGoxx.CH3Cl,16WaPaMo.CH3Cl,06NaBeBo.CH3Cl,11BrChMa.CH3Cl,13BrVoSc.CH3Cl,13SaLiMa.CH3Cl}.

 The HITRAN database~\cite{HITRAN} has the most detailed coverage with over $212{\,}000$ lines for the two main isotopologues, $^{12}$CH$_{3}{}^{35}$Cl and $^{12}$CH$_{3}{}^{37}$Cl (henceforth labelled as CH$_{3}{}^{35}$Cl and CH$_{3}{}^{37}$Cl). This includes rovibrational transitions up to $J=82$ and covers the $0${--}$3200{\,}$cm$^{-1}$ region. However, there are deficiencies and we will see in Sec.~\ref{sec:overview} that HITRAN is missing a band around $2880{\,}$cm$^{-1}$. Some line positions and intensities are also from theoretical predictions using a fairly old, empirically refined anharmonic force field~\citep{85KoKoNa.CH3Cl}. Given the numerous high-resolution studies since then, notably in the $3.4{\,}\mu$m region~\citep{11BrPeJa.CH3Cl} (included in HITRAN2012) and in the $6.9{\,}\mu$m region~\citep{13RaJaDh.CH3Cl}, improvements can be expected in the coverage of CH$_3$Cl. 
 Another valuable resource is the PNNL spectral library~\citep{PNNL} which covers the $600$ to $6500\,$cm$^{-1}$ region at a resolution of around $0.06\,$cm$^{-1}$ for temperatures of 5, 25 and 50$\,^{\circ}$C. Other databases such as GEISA~\citep{GEISA} include CH$_3$Cl but the datasets are not as extensive, whilst the JPL~\citep{JPL} catalog has been incorporated into HITRAN.

 Intensity information is vital for practical applications such as atmospheric modelling or remote sensing. The six fundamental bands of CH$_3$Cl have all been considered at some stage~\citep{11BrPeJa.CH3Cl,13RaJaDh.CH3Cl,77MaToxx.CH3Cl,78Maxxxx.CH3Cl,81DaMoGr.CH3Cl,88DaBlWa.CH3Cl,84ElKaSa.CH3Cl,91CaReTa.CH3Cl,89BlWaDa_1.CH3Cl,93BlWaMa.CH3Cl,93BlWaDa_1.CH3Cl,94BoBlWa.CH3Cl,94BlCoWa.CH3Cl,95BlWaPo.CH3Cl,95BlLaWa.CH3Cl,96BlLaWa.CH3Cl}. Notably, absolute line intensities have been measured for the $\nu_1$, $\nu_4$ and $3\nu_6$ bands in the $2920$ to $3100{\,}$cm$^{-1}$ range~\citep{11BrPeJa.CH3Cl}, and for over 900 rovibrational transitions in the $\nu_{5}$ band~\citep{13RaJaDh.CH3Cl}. These two studies are the most reliable and complete line intensity measurements for both CH$_{3}{}^{35}$Cl and CH$_{3}{}^{37}$Cl to date. From a theoretical perspective, calculations of dipole moment derivatives and infrared intensities have been reported~\citep{79Wixxxx.CH3Cl,83KoKoNa.CH3Cl.1,84KoKoNa.CH3Cl.1,85KoKoNa.CH3Cl,87ScThxx.CH3Cl,92ChPaxx.CH3Cl,95PaPaCh.CH3Cl,07JaTrXi.CH3Cl,16WaPaMo.CH3Cl}. However, we are unaware of any global DMS which could be used for intensity simulations of the rotation-vibration spectrum of CH$_3$Cl.

 Previously we reported~\citep{15OwYuYa.CH3Cl} two state-of-the-art \textit{ab initio} potential energy surfaces for the two main isotopologues of methyl chloride. Variational calculations of the vibrational $J=0$ energies and equilibrium geometry showed excellent agreement with experimental results. Building on this work, we present a new nine-dimensional \textit{ab initio} DMS which has been computed using high-level electronic structure theory. A symmetrized molecular bond (SMB) representation for XY$_3$Z-type molecules has been implemented into the nuclear motion code TROVE~\citep{TROVE2007} to represent the DMS analytically. Comprehensive calculations of the rotation-vibration spectrum are then carried out to evaluate the quality of the DMS. The work presented here represents the next step towards generating a complete rovibrational line list applicable for elevated temperatures.

 The paper is structured as follows: In Sec.~\ref{sec:DMS} the electronic structure calculations and analytic representation of the DMS are described. Details of the variational calculations are given in Sec.~\ref{sec:variational}. In Sec.~\ref{sec:results} the DMS is evaluated against a range of experimental measurements as well as the HITRAN and PNNL spectroscopic databases. Results include vibrational transition moments, absolute line intensities of the $\nu_1$, $\nu_4$, $\nu_5$ and $3\nu_6$ bands, and an overview of the rotation-vibration spectrum for states up to $J=85$ in the $0${--}$6500\,$cm$^{-1}$ frequency range. Concluding remarks are offered in Sec.~\ref{sec:conc}.

\section{Dipole Moment Surface}
\label{sec:DMS}

\subsection{Electronic structure calculations}

 The first derivative of the electronic energy with respect to external electric field strength defines the electric dipole moment of a molecule. Working in a Cartesian laboratory-fixed $XYZ$ coordinate system with origin at the C nucleus, an external electric field with components $\pm0.005{\,}$a.u. was applied along each axis and the respective dipole moment component $\mu_A$ for $A=X,Y,Z$ determined using finite differences. Calculations were carried out at the CCSD(T) [coupled cluster with all single and double excitations and a perturbational estimate of connected triple excitations] level of theory with the augmented correlation consistent quadruple zeta basis set, aug-cc-pVQZ(+d for Cl)~\cite{Dunning89,Kendall92,Woon93,Dunning01}, in the frozen core approximation. MOLPRO2012~\cite{Werner2012} was used for all calculations.

 The DMS was evaluated on a nine-dimensional global grid of $44{,}820$ points with energies up to $h c \cdot 50{\,}000{\,}$cm$^{-1}$ ($h$ is the Planck constant and $c$ is the speed of light). The grid included geometries in the range $1.3\leq r_0 \leq 2.95{\,}\mathrm{\AA}$, $0.7\leq r_i \leq 2.45{\,}\mathrm{\AA}$, $65\leq \beta_i \leq 165^{\circ}$ for $i=1,2,3$ and $55\leq \tau_{jk} \leq 185^{\circ}$ with $jk=12,13$. Here, the nine internal coordinates are: the C--Cl bond length $r_0$; three C--H bond lengths $r_1$, $r_2$ and $r_3$; three $\angle(\mathrm{H}_i\mathrm{CCl})$ interbond angles $\beta_1$, $\beta_2$ and $\beta_3$; and two dihedral angles $\tau_{12}$ and $\tau_{13}$ between adjacent planes containing H$_i$CCl and H$_j$CCl. The grid utilized for the DMS is the same as that used for the PESs previously reported~\citep{15OwYuYa.CH3Cl}.

\subsection{Analytic representation}

 Before fitting an analytic expression to the \textit{ab initio} data it is necessary to establish a suitable molecule-fixed $xyz$ coordinate system. Methyl chloride is a prolate symmetric top molecule of the $\bm{C}_{3\mathrm{v}}\mathrm{(M)}$ symmetry group~\cite{MolSym_BuJe98}. There are six symmetry operations $\left\lbrace E,(123),(132),(12)^{*},(23)^{*},(13)^{*}\right\rbrace$ which make up $\bm{C}_{3\mathrm{v}}$(M). The cyclic permutation operation $(123)$ replaces nucleus 1 with nucleus 2, nucleus 2 with nucleus 3, and nucleus 3 with nucleus 1; the permutation-inversion operation $(12)^{*}$ interchanges nuclei 1 and 2 and inverts all particles (including electrons) in the molecular centre of mass; the identity operation $E$ leaves the molecule unchanged. The symmetrized molecular bond (SMB) representation has been successfully applied to molecules of $\bm{C}_{3\mathrm{v}}$(M) symmetry~\cite{YuBaYa09.NH3,UnTeYu13.SO3} and this approach is employed for the present study.

 We first define unit vectors along each of the four bonds of CH$_3$Cl,
\begin{equation}
\mathbf{e}_i = \frac{\mathbf{r}_i-\mathbf{r}_{\mathrm{C}}}{\lvert \mathbf{r}_i-\mathbf{r}_{\mathrm{C}}\rvert}{\,};\hspace{4mm}i=0,1,2,3 ,
\end{equation}
where $\mathbf{r}_{\mathrm{C}}$ is the position vector of the C nucleus, $\mathbf{r}_0$ the Cl nucleus, and $\mathbf{r}_1$, $\mathbf{r}_2$ and $\mathbf{r}_3$ the respective H atoms. The \textit{ab initio} dipole moment vector ${\bm\mu}$ is projected onto the molecular bonds and can be described by molecule-fixed $xyz$ dipole moment components,
\begin{align}
\mu_x &= \frac{1}{\sqrt{6}}\left(2({\bm\mu}\cdot\mathbf{e}_1) - ({\bm\mu}\cdot\mathbf{e}_2) - ({\bm\mu}\cdot\mathbf{e}_3) \right) , \label{eq:mu_x}\\
\mu_y &= \frac{1}{\sqrt{2}}\left(({\bm\mu}\cdot\mathbf{e}_2) - ({\bm\mu}\cdot\mathbf{e}_3) \right) , \label{eq:mu_y}\\
\mu_z &= {\bm\mu}\cdot\mathbf{e}_0 .\label{eq:mu_z}
\end{align}
Symmetry-adapted combinations have been formed for $\mu_{x}$ and $\mu_{y}$ and these transform according to $E$ symmetry, while the $\mu_{z}$ component is of $A_1$ symmetry. The advantage of the SMB representation is that the unit vectors $\mathbf{e}_i$ used to define $\mathbf{\mu}$ for any  instantaneous positions of the nuclei are related to the internal coordinates only.

 To construct the three dipole surfaces corresponding to the components given in Eqs.~\eqref{eq:mu_x} to \eqref{eq:mu_z}, a numerical, on-the-fly symmetrization procedure has been implemented. This is similar to the approach employed for the PES~\citep{15OwYuYa.CH3Cl} but because ${\bm\mu}$ is a vector quantity we have to consider the transformation properties of the dipole moment components themselves. For $\mu_z$, which points along the C--Cl bond, the process is trivial owing to its $A_1$ symmetry and invariance to the $\bm{C}_{3\mathrm{v}}$(M) symmetry operations. Building an analytic expression follows the same steps as the PES. For the two $E$ symmetry components, $\mu_x$ and $\mu_y$, the construction is more subtle and they must be treated together.

 We consider an initial (reference) term in the dipole expansion belonging to $\mu_x$,
\begin{equation}\label{eq:mux}
\left( \begin{matrix} \mu_x \\ \mu_y \end{matrix} \right) = \left( \begin{matrix} \mu_{x,ijk\ldots}^{\mathrm{initial}} \\ 0 \end{matrix} \right) ,
\end{equation}
where
\begin{equation}
\label{eq:mu_initial}
\mu_{x,ijk\ldots}^{\mathrm{initial}}=\left(\xi_{1}^{\,i}\xi_{2}^{\,j}\xi_{3}^{\,k}\xi_{4}^{\,l}\xi_{5}^{\,m}\xi_{6}^{\,n}\xi_{7}^{\,p}\xi_{8}^{\,q}\xi_{9}^{\,r}\right) .
\end{equation}
This term has maximum expansion order $i+j+k+l+m+n+p+q+r=6$, and is expressed in terms of the nine coordinates,
\begin{align}
\xi_1&=\left(r_0 - r_0^{\mathrm{ref}}\right) ,\label{eq:stretch1}\\
\xi_j&=\left(r_i - r_1^{\mathrm{ref}}\right){\,};\hspace{2mm}j=2,3,4{\,}, \hspace{2mm} i=j-1 ,\label{eq:stretch2}\\
\xi_k &= (\beta_i - \beta^{\mathrm{ref}}){\,};\hspace{2mm}k=5,6,7{\,}, \hspace{2mm} i=k-4 ,\label{eq:angular1}\\
\xi_8 &= \frac{1}{\sqrt{6}}\left(2\tau_{23}-\tau_{13}-\tau_{12}\right) ,\label{eq:angular2}\\
\xi_9 &= \frac{1}{\sqrt{2}}\left(\tau_{13}-\tau_{12}\right) .\label{eq:angular3}
\end{align}
Here, $\tau_{23}=2\pi-\tau_{12}-\tau_{13}$ and the reference structural parameters $r_0^{\mathrm{ref}}=1.7550{\,}\mathrm{\AA}$, $r_1^{\mathrm{ref}}=1.0415{\,}\mathrm{\AA}$ and $\beta^{\mathrm{ref}}=108.414^{\circ}$. Note that the values of $r_0^{\mathrm{ref}}$, $r_1^{\mathrm{ref}}$ and $\beta^{\mathrm{ref}}$ were optimized during the fitting of the DMS.

 The action of a symmetry operation $\mathbf{X}=\lbrace E,(123),(132),(12)^{*},(23)^{*},(13)^{*}\rbrace$ on Eq.~\eqref{eq:mu_initial} will (i) permute the expansion indices $ijk\ldots$, to $i\p j\p k\p\ldots$ to produce a new expansion term and (ii) permute the unit vectors $\mathbf{e}_i$ for $i=1,2,3$. Using the projection operator technique~\citep{MolSym_BuJe98}, this latter contribution is projected onto the $\mathbf{e}_x$ and $\mathbf{e}_y$ molecule-fixed vectors and added to the respective dipole moment components. The resulting components, $\mu^{\prime}_x$ and $\mu^{\prime}_y$, reduce to
\begin{equation}
\left( \begin{matrix} \mu^{\prime}_x \\ \mu^{\prime}_y \end{matrix} \right) = \left( \begin{matrix} C_1\mu_{x,ijk\ldots}^{\mathbf{X}} \\ C_2\mu_{x,ijk\ldots}^{\mathbf{X}} \end{matrix} \right) ,
\end{equation}
where $C_1$ and $C_2$ are constants associated with the acting $\bm{C}_{3\mathrm{v}}$(M) symmetry operation, and $\mu_{x,ijk\ldots}^{\mathbf{X}}$ is the new expansion term connected to Eq.~\eqref{eq:mu_initial} by the symmetry operation $\mathbf{X}$. Note that a contribution arises in $\mu^{\prime}_y$ ($C_2\ne 0$) due to the projection operator acting on the two-component quantity $(\mu_x,\mu_y)$.

 The steps are repeated for each symmetry operation of $\bm{C}_{3\mathrm{v}}$(M) and the results summed to produce a final dipole term (ignoring constants),
\begin{equation}\label{eq:dip_final}
\mu^{\mathrm{final}}_{x,ijk\ldots}=\mu_{x,ijk\ldots}^{E}+\mu_{x,ijk\ldots}^{(123)}+\mu_{x,ijk\ldots}^{(132)}+\mu_{x,ijk\ldots}^{(12)^{*}}+\mu_{x,ijk\ldots}^{(23)^{*}}+\mu_{x,ijk\ldots}^{(13)^{*}} ,
\end{equation}
which is best understood as a sum of symmetrized combinations of different permutations of coordinates $\xi_{i}$. Likewise, a similar expression contributes to $\mu_y$. Although we have only considered an initial term belonging to $\mu_x$, the same idea applies to initial terms belonging to $\mu_y$. Incorporating $\mu_z$ into the procedure is straightforward, thus enabling the simultaneous construction of all three dipole moment surfaces of CH$_3$Cl. Each surface is represented by the analytic expression
\begin{equation}\label{eq:mu_tot}
\mu_{\alpha}^{\mathrm{total}}(\xi_{1},\xi_{2},\xi_{3},\xi_{4},\xi_{5},\xi_{6},\xi_{7},\xi_{8},\xi_{9})={\sum_{ijk\ldots}}F^{(\alpha)}_{ijk\ldots}\mu^{\mathrm{final}}_{\alpha,ijk\ldots} ,
\end{equation}
where some of the expansion coefficients $F^{(\alpha)}_{ijk\ldots}$ are shared between the $x$ and $y$ components. 

 A least squares fitting to the \textit{ab initio} data utilizing Watson's robust fitting scheme~\citep{Watson03} was employed to determine $F^{(\alpha)}_{ijk\ldots}$ for $\alpha=x,y,z$. Weight factors of the form~\citep{Schwenke97},
\begin{equation}\label{eq:weights}
w_i=\left(\frac{\tanh\left[-0.0006\times(\tilde{E}_i - 15{\,}000)\right]+1.002002002}{2.002002002}\right)\times\frac{1}{N\tilde{E}_i^{(w)}} ,
\end{equation}
were used in the fitting, with normalization constant $N=0.0001$ and $\tilde{E}_i^{(w)}=\max(\tilde{E}_i, 10{\,}000)$, where $\tilde{E}_i$ is the potential energy at the $i$th geometry above equilibrium (all values in cm$^{-1}$). At geometries where $r_0\geq 2.35{\,}\mathrm{\AA}$, or $r_i\geq 2.00{\,}\mathrm{\AA}$ for $i=1,2,3$, the weights were decreased by several orders of magnitude. This was done because the coupled cluster method is known to become unreliable at very large stretch coordinates, indicated by a T1 diagnostic value $>0.02$~\cite{T1_Lee89}. Whilst the energies are not wholly accurate at these points, they ensure the DMS maintains a reasonable shape towards dissociation.

 The three dipole surfaces for $\mu_x$, $\mu_y$ and $\mu_z$ employed sixth order expansions and used 175, 163 and 235 parameters, respectively. A combined weighted root-mean-square (rms) error of $9\times10^{-5}{\,}$D was obtained for the fitting. Incorporating the analytic representation into variational nuclear motion calculations is relatively straightforward and the implementation requires only a small amount of code. The dipole expansion parameters along with a FORTRAN routine to construct the DMS are provided in the supplementary material.

\section{Variational calculations}
\label{sec:variational}

 The nuclear motion program TROVE~\cite{TROVE2007} was employed for all rovibrational calculations and details of the general methodology can be found elsewhere~\citep{TROVE2007,YuBaYa09.NH3,15YaYu.ADF}. Since methyl chloride has already been treated using TROVE~\citep{15OwYuYa.CH3Cl}, we summarize only the key aspects of our calculations.

 An automatic differentiation method~\cite{15YaYu.ADF} was used to construct the rovibrational Hamiltonian numerically. The Hamiltonian itself was represented as a power series expansion around the equilibrium geometry in terms of nine vibrational coordinates. The coordinates used are almost identical to those given in Eqs.~\eqref{eq:stretch1} to \eqref{eq:angular3}, except for the potential energy operator where Morse oscillator functions replace the linear expansion variables for stretching modes. In all calculations the kinetic and potential energy operators were truncated at 6th and 8th order, respectively. This level of truncation is adequate for our purposes (see Ref.~\citep{TROVE2007} and \citep{15YaYu.ADF} for a discussion of the associated errors of such a scheme). Atomic mass values were employed throughout.

 Two purely \textit{ab initio} PESs~\citep{15OwYuYa.CH3Cl}, CBS-35$^{\,\mathrm{HL}}$ and CBS-37$^{\,\mathrm{HL}}$, corresponding to the two main isotopologues, CH$_{3}{}^{35}$Cl and CH$_{3}{}^{37}$Cl, have been utilized for the present study. The surfaces are based on extensive explicitly correlated coupled cluster calculations with extrapolation to the complete basis set (CBS) limit and include a range of additional higher-level energy corrections. The CBS-35$^{\,\mathrm{HL}}$ and CBS-37$^{\,\mathrm{HL}}$ PESs reproduce the fundamental term values with rms errors of $0.75$ and $1.00{\,}$cm$^{-1}$, respectively. We are therefore confident that the DMS of CH$_3$Cl can be evaluated accurately in conjunction with these PESs.

 A multi-step contraction scheme \cite{YuBaYa09.NH3} is used to construct the vibrational basis set and a polyad number truncation scheme controls its size. For CH$_3$Cl, we define the polyad number
\begin{equation}\label{eq:polyad}
P = n_1+2(n_2+n_3+n_4)+n_5+n_6+n_7+n_8+n_9 \leq P_{\mathrm{max}} ,
\end{equation}
and this does not exceed a predefined maximum value $P_{\mathrm{max}}$. Here, the quantum numbers $n_k$ for $k=1,\ldots,9$ correspond to primitive basis functions $\phi_{n_k}$ for each vibrational mode. Multiplication with rigid-rotor eigenfunctions $\ket{J,K,m,\tau_{\mathrm{rot}}}$ produces the final symmetrized basis set for use in $J>0$ calculations. The quantum number $K$ is the projection (in units of $\hbar$) of $J$ onto the molecule-fixed $z$-axis, whilst $\tau_{\mathrm{rot}}$ determines the rotational parity as $(-1)^{\tau_{\mathrm{rot}}}$. As we will see in Sec.~\ref{sec:results}, different sized basis sets have been utilized in this work and this reflects the computational demands of variational calculations of rovibrational spectra.

 TROVE automatically assigns quantum numbers to the eigenvalues and corresponding eigenvectors by analysing the contribution of the basis functions. To be of spectroscopic use we map the vibrational quantum numbers $n_k$ to the normal mode quantum numbers $\mathrm{v}_k$ commonly used in spectroscopic studies. For CH$_3$Cl, vibrational states are labelled as $\mathrm{v_1}\nu_1+\mathrm{v_2}\nu_2+\mathrm{v_3}\nu_3+\mathrm{v_4}\nu_4+\mathrm{v_5}\nu_5+\mathrm{v_6}\nu_6$ where $\mathrm{v_i}$ counts the level of excitation.

  The normal modes of methyl chloride are of $A_{1}$ or $E$ symmetry. The three non-degenerate modes have $A_{1}$ symmetry; the symmetric CH$_{3}$ stretching mode $\nu_{1}$ ($2967.77/2967.75{\,}$cm$^{-1}$), the symmetric CH$_{3}$ deformation mode $\nu_{2}$ ($1354.88/1354.69{\,}$cm$^{-1}$) and the C{--}Cl stretching mode $\nu_{3}$ ($732.84/727.03{\,}$cm$^{-1}$). Whilst the three degenerate modes have $E$ symmetry; the CH$_{3}$ stretching mode $\nu_{4}^{\ell_{4}}$ ($3039.26/3039.63{\,}$cm$^{-1}$), the CH$_{3}$ deformation mode $\nu_{5}^{\ell_{5}}$ ($1452.18/1452.16{\,}$cm$^{-1}$) and the CH$_{3}$ rocking mode $\nu_{6}^{\ell_{6}}$ ($1018.07/1017.68{\,}$cm$^{-1}$). The values in parentheses are the experimentally determined fundamental frequencies for CH$_{3}{}^{35}$Cl / CH$_{3}{}^{37}$Cl~\citep{11BrPeJa.CH3Cl,05NiChBu.CH3Cl}. The additional vibrational angular momentum quantum numbers $\ell_{4}$, $\ell_{5}$ and $\ell_{6}$ are necessary to resolve the degeneracy of their respective modes.

\section{Results}
\label{sec:results}

\subsection{Vibrational transition moments}
\label{sec:TM}

 As an initial test of the DMS we compute vibrational transition moments,
\begin{equation}
\mu_{if} = \sqrt{\sum_{\alpha=x,y,z}{\lvert\bra{\Phi^{(f)}_{\mathrm{vib}}}\bar{\mu}_{\alpha}\ket{\Phi^{(i)}_{\mathrm{vib}}}\rvert}^2} .
\end{equation}
Here, $\ket{\Phi^{(i)}_{\mathrm{vib}}}$ and $\ket{\Phi^{(f)}_{\mathrm{vib}}}$ are the initial and final state vibrational $(J=0)$ eigenfunctions, respectively, and $\bar{\mu}_{\alpha}$ is the electronically averaged dipole moment function along the molecule-fixed axis $\alpha=x,y,z$.

 Transition moments have been determined experimentally for the six fundamental modes of CH$_3{}^{35}$Cl and these are listed in Table~\ref{tab:TM} along with our computed values. Calculations employed a polyad truncation number of $P_{\mathrm{max}}=12$ which is sufficient for converging $\mu_{if}$. Overall the agreement is encouraging and it indicates that the DMS should be reliable for intensity simulations of the fundamental bands.

\begin{table}
\begin{center}
\tabcolsep=0.15cm
\caption{\label{tab:TM}Calculated vibrational transition moments (in Debye) and frequencies (in cm$^{-1}$) from the vibrational ground state for CH$_3{}^{35}$Cl and CH$_3{}^{37}$Cl.}
\resizebox{\linewidth}{!}{
\begin{tabular}{c c c c c l l}
\hline\hline
Mode & Sym. & Experiment$^a$ & Calculated & \multicolumn{1}{c}{$\mu_{if}^{\mathrm{calc}}$} & \multicolumn{1}{c}{$\mu_{if}^{\mathrm{exp}}$} & \multicolumn{1}{c}{Ref.}\\
\hline
& & & & CH$_3{}^{35}$Cl & & \\[-1mm] 
$\nu_1$ & $A_1$ & 2967.77 & 2969.16 & 0.05296 & 0.053$^b$ & \citet{84ElKaSa.CH3Cl}\\[-1mm]
$\nu_2$ & $A_1$ & 1354.88 & 1355.01 & 0.05260 & 0.05006(1)$^c$ & \citet{95BlLaWa.CH3Cl}\\[-1mm]
$\nu_3$ & $A_1$ & 732.84 & 733.22 & 0.11468 & 0.1121(8)  & \citet{88DaBlWa.CH3Cl}\\[-1mm]
$\nu_4$ & $E$ & 3039.26 & 3038.19 & 0.03108 & 0.033$^d$ & \citet{84ElKaSa.CH3Cl}\\[-1mm]
$\nu_5$ & $E$ & 1452.18 & 1452.56 & 0.05451 & 0.0527(7) & \citet{91CaReTa.CH3Cl}\\[-1mm]
$\nu_6$ & $E$ & 1018.07 & 1018.05 & 0.03707 & 0.0388$^e$  & \citet{93BlWaMa.CH3Cl}\\
& & & & CH$_3{}^{37}$Cl & & \\[-1mm] 
$\nu_1$ & $A_1$ & 2967.75 & 2969.14 & 0.05296 & \multicolumn{1}{c}{{--}} & \multicolumn{1}{c}{{--}}\\[-1mm]
$\nu_2$ & $A_1$ & 1354.69 & 1354.82 & 0.05275 & \multicolumn{1}{c}{{--}} & \multicolumn{1}{c}{{--}}\\[-1mm]
$\nu_3$ & $A_1$ & 727.03 & 727.40 & 0.11416 & \multicolumn{1}{c}{{--}}  & \multicolumn{1}{c}{{--}}\\[-1mm]
$\nu_4$ & $E$ & 3039.63 & 3037.71 & 0.02939 & \multicolumn{1}{c}{{--}} & \multicolumn{1}{c}{{--}}\\[-1mm]
$\nu_5$ & $E$ & 1452.16 & 1452.53 & 0.05449 & \multicolumn{1}{c}{{--}} & \multicolumn{1}{c}{{--}}\\[-1mm]
$\nu_6$ & $E$ & 1017.68 & 1017.66 & 0.03724 & \multicolumn{1}{c}{{--}} & \multicolumn{1}{c}{{--}}\\
\hline\hline
\end{tabular}}
\end{center}
{\footnotesize $^a$ From \citet{11BrPeJa.CH3Cl} and \citet{05NiChBu.CH3Cl}. $^b$ From \citet{95PaPaCh.CH3Cl} but derived from band strength measurement of $S_{\mathrm{v}}=84.3\pm3.3\,$cm$^{-2}\,$atm$^{-1}$ at $296\,$K~\citep{84ElKaSa.CH3Cl}. $^c$ Value of $\mu_{\nu_2}^{\mathrm{exp}}=0.0473(7)\,$D determined in \citet{91CaReTa.CH3Cl}. $^d$ From \citet{95PaPaCh.CH3Cl} but derived from band strength measurement of $S_{\mathrm{v}}=33.6\pm1.4\,$cm$^{-2}\,$atm$^{-1}$at $296\,$K~\citep{84ElKaSa.CH3Cl}. $^e$ From \citet{95PaPaCh.CH3Cl} but derived from band strength measurement of $S_{\mathrm{v}}=15.1\pm1.6\,$cm$^{-2}\,$atm$^{-1}$at $296\,$K~\citep{93BlWaMa.CH3Cl}.}
\end{table}

 For CH$_3{}^{37}$Cl, band strength measurements of the $\nu_3$~\citep{89BlWaDa_1.CH3Cl} and $\nu_6$~\citep{93BlWaDa_1.CH3Cl} bands have been carried out but only minor differences were observed compared to CH$_3{}^{35}$Cl~\citep{88DaBlWa.CH3Cl,93BlWaMa.CH3Cl}. Likewise, as seen in Table~\ref{tab:TM} the computed transition moments for the fundamentals only marginally differ compared to CH$_3{}^{35}$Cl. It seems the intensity variation from isotopic substitution in methyl chloride is relatively small and in some instances almost negligible. A list of computed transitions moments from the vibrational ground state for $79$ levels up to $4200\,$cm$^{-1}$ is provided in the supplementary material. Note that for the equilibrium dipole moment of methyl chloride we calculate $\mu=1.8909\,$D which is close to the experimental value of $\mu=1.8959(15)\,$D~\citep{85WlSeLe.CH3Cl}.

\subsection{Absolute line intensities of the $\nu_1$, $\nu_4$, $\nu_5$ and $3\nu_6$ bands}
\label{sec:abs_intens}

 Recently, absolute line intensities were determined for the $\nu_1$, $\nu_4$ and $3\nu_6$ bands around the $3.4{\,}\mu$m region~\citep{11BrPeJa.CH3Cl} (included in HITRAN2012), and for the $\nu_5$ band in the $6.9{\,}\mu$m region~\citep{13RaJaDh.CH3Cl}. To assess the DMS we compare against these works for both isotopologues up to $J=15$. Calculating higher rotational excitation is computationally demanding (rovibrational matrices scale linearly with $J$) so we set $P_{\mathrm{max}}=10$ which is sufficient for reliable intensities. A study on the five-atom molecule SiH$_4$~\citep{15OwYuYa.SiH4}, which has similar convergence properties with respect to $P_{\mathrm{max}}$, also employed $P_{\mathrm{max}}=10$ to produce intensities of the $\nu_3$ band with an estimated convergence error of $1\%$ or less for transitions up to $J=16$. Because the two \textit{ab initio} PESs used in this study, CBS-35$^{\,\mathrm{HL}}$ and CBS-37$^{\,\mathrm{HL}}$, can at best only be considered accurate to about $\pm 1\,$cm$^{-1}$, for illustrative purposes we have shifted computed line positions to better match experiment in the following comparisons.

 Absolute absorption intensities have been simulated at room temperature ($T=296\,$K) using the expression,
\begin{equation}
I(f \leftarrow i) = \frac{A_{if}}{8\pi c}g_{\mathrm{ns}}(2 J_{f}+1)\frac{\exp\left(-E_{i}/kT\right)}{Q(T)\; \nu_{if}^{2}}\left[1-\exp\left(-\frac{hc\nu_{if}}{kT}\right)\right] ,
\end{equation}
where $A_{if}$ is the Einstein $A$ coefficient of a transition with frequency $\nu_{if}$ between an initial state with energy $E_i$, and a final state with rotational quantum number $J_f$. Here, $k$ is the Boltzmann constant, $T$ is the absolute temperature and $c$ is the speed of light. The nuclear spin statistical weights are $g_{\mathrm{ns}}=\lbrace 16,16,16\rbrace$ for states of symmetry $\lbrace A_1,A_2,E\rbrace$, respectively. These values have been calculated using the method detailed in \citet{gns:Jensen1999}. For the partition function we use values of $Q(T)=57,915.728$ and $58,833.711$ for CH$_3{}^{35}$Cl and CH$_3{}^{37}$Cl, respectively~\citep{HITRAN}.  Note that to ensure a correct comparison with the experimental studies of \citet{11BrPeJa.CH3Cl} and \citet{13RaJaDh.CH3Cl}, the intensities of overlapping $A_1$ and $A_2$ spectral lines (listed as being of $A$ symmetry) must be halved.

\begin{figure}
\includegraphics[width=\textwidth,angle=0]{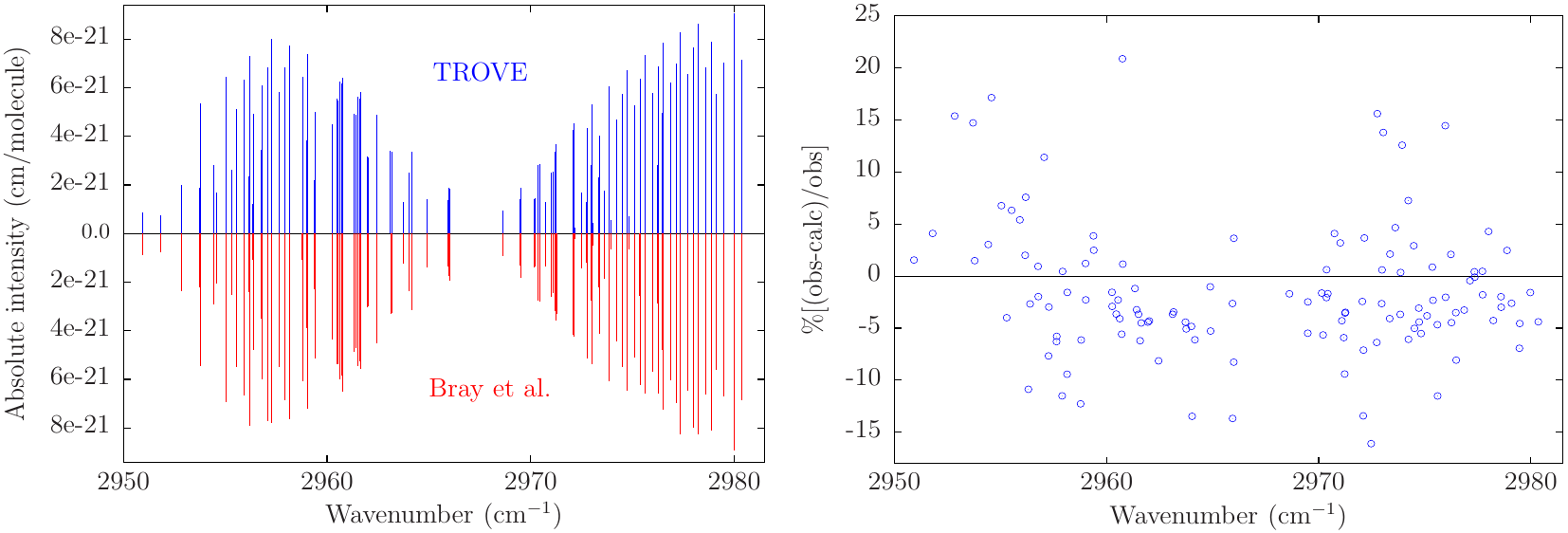}
\caption{\label{fig:v1_band}Absolute line intensities of the $\nu_1$ band for transitions up to $J=15$ (left) and the corresponding residuals $\left(\%\left[\frac{\mathrm{obs-calc}}{\mathrm{obs}}\right]\right)$ (right) when compared with measurements from \citet{11BrPeJa.CH3Cl}. Transitions for both CH$_3{}^{35}$Cl and CH$_3{}^{37}$Cl are shown and the intensities have not been scaled to natural abundance. For illustrative purposes TROVE line positions have been shifted by $-1.35{\,}$cm$^{-1}$.}
\end{figure}

\begin{figure}
\includegraphics[width=\textwidth,angle=0]{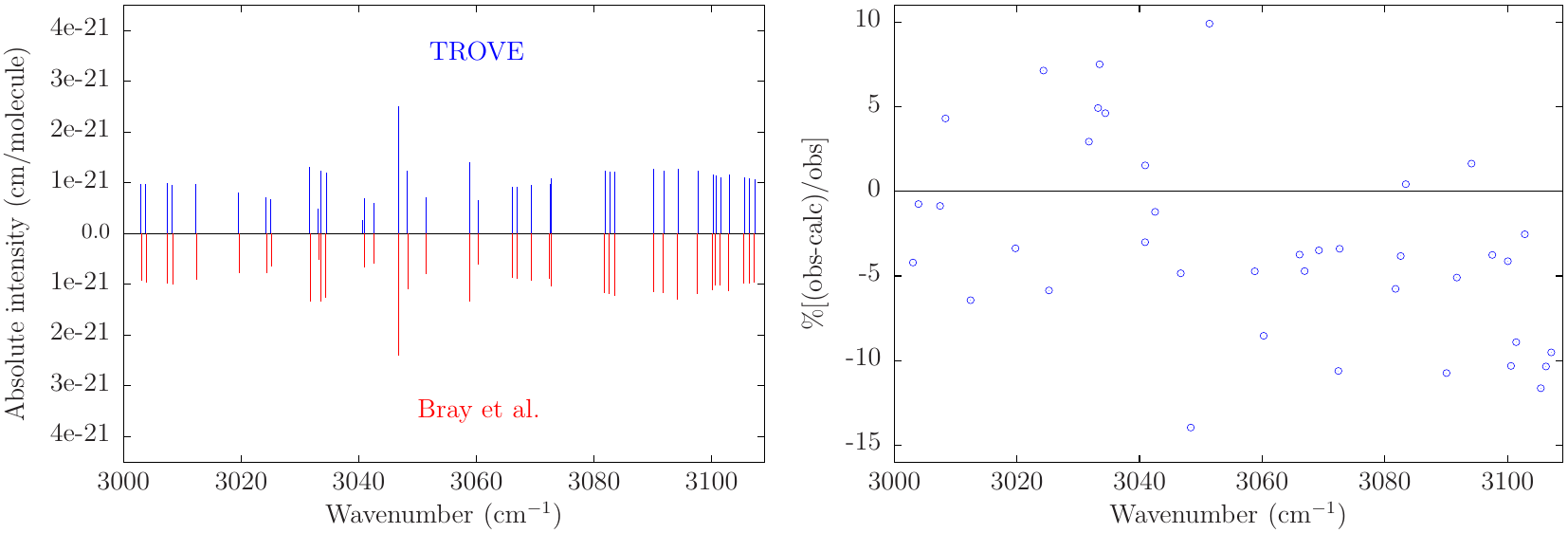}
\caption{\label{fig:v4_band}Absolute line intensities of the $\nu_4$ band for transitions up to $J=15$ (left) and the corresponding residuals $\left(\%\left[\frac{\mathrm{obs-calc}}{\mathrm{obs}}\right]\right)$ (right) when compared with measurements from \citet{11BrPeJa.CH3Cl}. Transitions for both CH$_3{}^{35}$Cl and CH$_3{}^{37}$Cl are shown and the intensities have not been scaled to natural abundance. For illustrative purposes TROVE line positions have been shifted by $-1.42{\,}$cm$^{-1}$.}
\end{figure}

 In Fig.~\ref{fig:v1_band} we plot absolute line intensities for $126$ transitions of the $\nu_1$ band and their corresponding residuals $\left(\%\left[\frac{\mathrm{obs-calc}}{\mathrm{obs}}\right]\right)$ compared to measurements from \citet{11BrPeJa.CH3Cl}. The majority of computed intensities, although tending to be marginally stronger, are within the experimental accuracy of $10\%$ or better~\citep{private_comm}. Calculated line positions had on average a residual error of $\Delta_{\mathrm{obs}-\mathrm{calc}}=-1.35{\,}$cm$^{-1}$ and this has been corrected for in Fig.~\ref{fig:v1_band}. Similarly, computed intensities of the $\nu_4$ band shown in Fig.~\ref{fig:v4_band} are largely within experimental uncertainty. Here, line positions possessed a residual error of $\Delta_{\mathrm{obs}-\mathrm{calc}}=-1.42{\,}$cm$^{-1}$.

\begin{figure}
\includegraphics[width=\textwidth,angle=0]{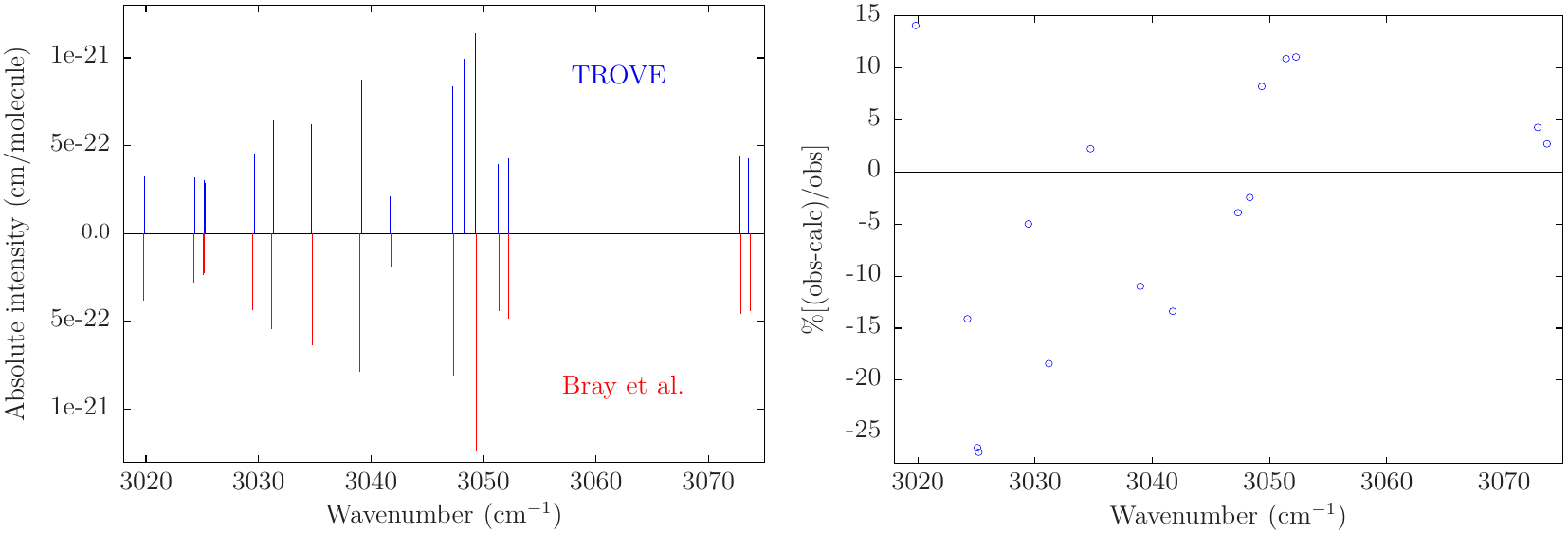}
\caption{\label{fig:3v6_band}Absolute line intensities of the $3\nu_6$ band for transitions up to $J=15$ (left) and the corresponding residuals $\left(\%\left[\frac{\mathrm{obs-calc}}{\mathrm{obs}}\right]\right)$ (right) when compared with measurements from \citet{11BrPeJa.CH3Cl}. Transitions for both CH$_3{}^{35}$Cl and CH$_3{}^{37}$Cl are shown and the intensities have not been scaled to natural abundance. For illustrative purposes TROVE line positions have been shifted by $-1.23{\,}$cm$^{-1}$.}
\end{figure}

 Line intensities of the $3\nu_6$ band are shown in Fig.~\ref{fig:3v6_band}. Excited modes are harder to converge and the size of the vibrational basis set at $P_{\mathrm{max}}=10$ means the respective rovibrational energy levels have a convergence error of $1.0\,$cm$^{-1}$ for low $J$ values (compared to errors of $\approx 0.1$, $0.5$ and $0.03\,$cm$^{-1}$ for the $\nu_1$, $\nu_4$ and $\nu_5$ bands, respectively). The effect is that computed line intensities will have an uncertainty of around $5\%$. Even so, the agreement for the $16$ lines from \citet{11BrPeJa.CH3Cl} is good. Note that line positions displayed a residual error of $\Delta_{\mathrm{obs}-\mathrm{calc}}=-1.23{\,}$cm$^{-1}$.
 
\begin{figure}
\includegraphics[width=\textwidth,angle=0]{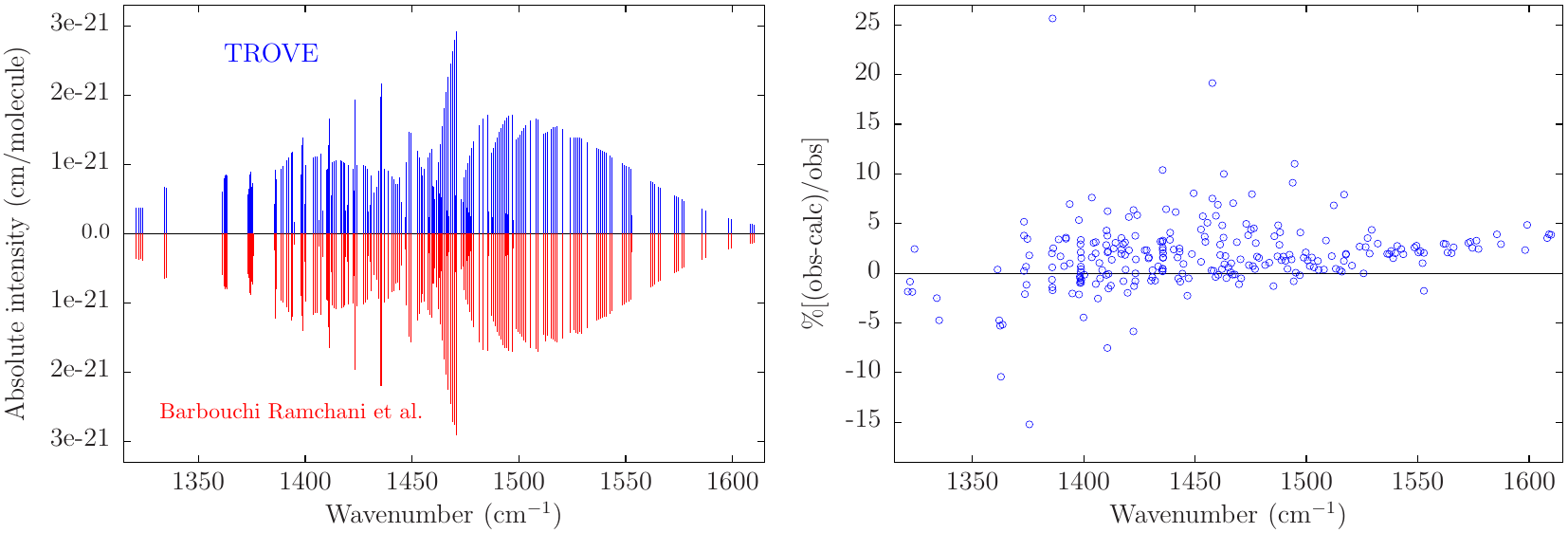}
\caption{\label{fig:v5_band}Absolute line intensities of the $\nu_5$ band for transitions up to $J=15$ (left) and the corresponding residuals $\left(\%\left[\frac{\mathrm{obs-calc}}{\mathrm{obs}}\right]\right)$ (right) when compared with measurements from \citet{13RaJaDh.CH3Cl}. Transitions for both CH$_3{}^{35}$Cl and CH$_3{}^{37}$Cl are shown and the intensities have not been scaled to natural abundance. For illustrative purposes TROVE line positions have been shifted by $-0.40{\,}$cm$^{-1}$.}
\end{figure}

 A high-resolution study of the $\nu_5$ band measured absolute line intensities with an experimental accuracy of $5\%$ or less, and line positions with an average estimated accuracy between $10^{-3}$ to $10^{-4}\,$cm$^{-1}$. As shown in Fig.~\ref{fig:v5_band} a significant number of computed line intensities are within experimental uncertainty and agreement for the $256$ transitions up to $J=15$ is excellent. Here calculated line positions had a residual error of $\Delta_{\mathrm{obs}-\mathrm{calc}}=-0.40{\,}$cm$^{-1}$.

\subsection{Overview of rotation-vibration line list}
\label{sec:overview}

 The HITRAN database contains over $212{\,}000$ lines for CH$_3$Cl and considers transitions up to $J=82$. To compute such highly excited rovibrational energy levels it has been necessary to again reduce the size of the vibrational basis set. Calculations were carried out with $P_{\mathrm{max}}=8$ and an upper energy level cut-off of $8000\,$cm$^{-1}$. Subsequent transitions and intensities were computed for a $6300\,$cm$^{-1}$ frequency window with a lower state energy threshold of $4400{\,}$cm$^{-1}$. Information has undoubtedly been lost by introducing these thresholds but the values were carefully chosen to keep this to a minimum. Such restrictions also allow the straightforward calculation of high $J$ values in a timely manner on compute nodes with $64\,$GB of RAM. Note that for pure rotational transitions in HITRAN the hyperfine structure has been resolved~\citep{A_HITRAN:2006}. Therefore, in order to have a reliable comparison for this spectral region we scale our intensities by a factor of $1/2$.

\begin{figure}
\includegraphics[width=\textwidth,angle=0]{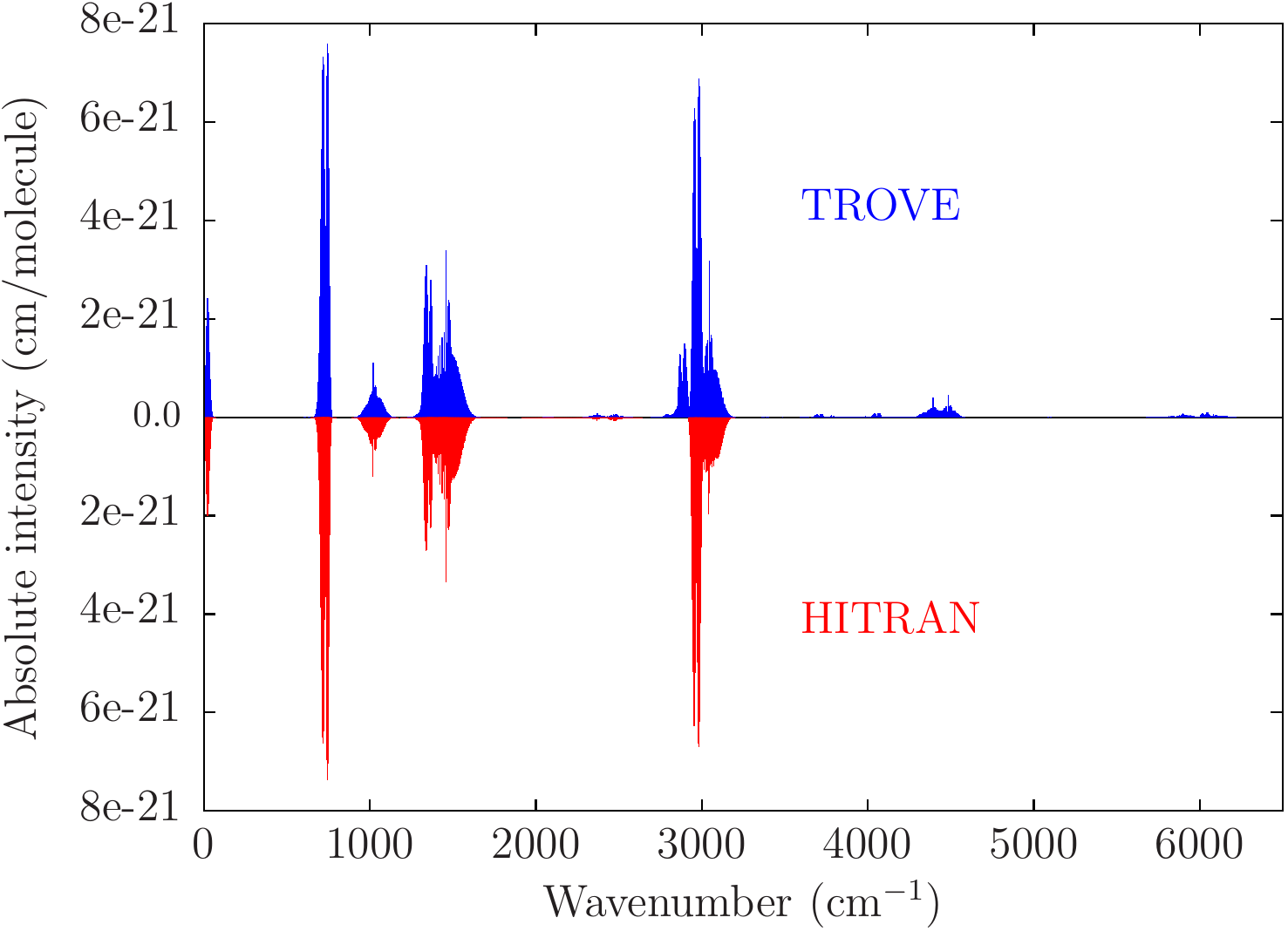}
\caption{\label{fig:hitran}Overview of methyl chloride rotation-vibration line list up to $J=85$ compared with all transitions in the HITRAN database~\citep{HITRAN}. Computed intensities have been scaled to natural abundance.}
\end{figure}

\begin{figure}
\centering
\includegraphics{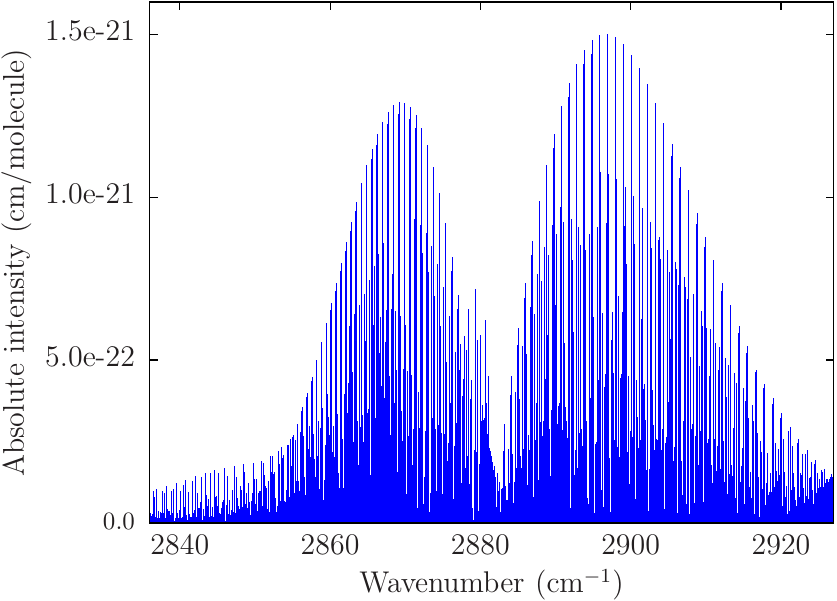}
\caption{\label{fig:2v5}The $2\nu_5$ band of methyl chloride. Computed intensities have been scaled to natural abundance.}
\end{figure}

 In Fig.~\ref{fig:hitran} we present a computed line list up to $J=85$ for both isotopologues of methyl chloride. Computed intensities have been scaled to natural abundance (0.748937 for CH$_3{}^{35}$Cl and 0.239491 for CH$_3{}^{37}$Cl) and are compared against all available lines in the HITRAN database. Overall the agreement is pleasing, particularly given the reduced size of the vibrational basis set and energy level thresholds. Up to $3200\,$cm$^{-1}$ the only noticeable missing band in HITRAN appears to be the $2\nu_5$ band around $2880\,$cm$^{-1}$shown in Fig.~\ref{fig:2v5}. This is not expected to be important for atmospheric sensing.

\begin{figure}
\centering
\includegraphics{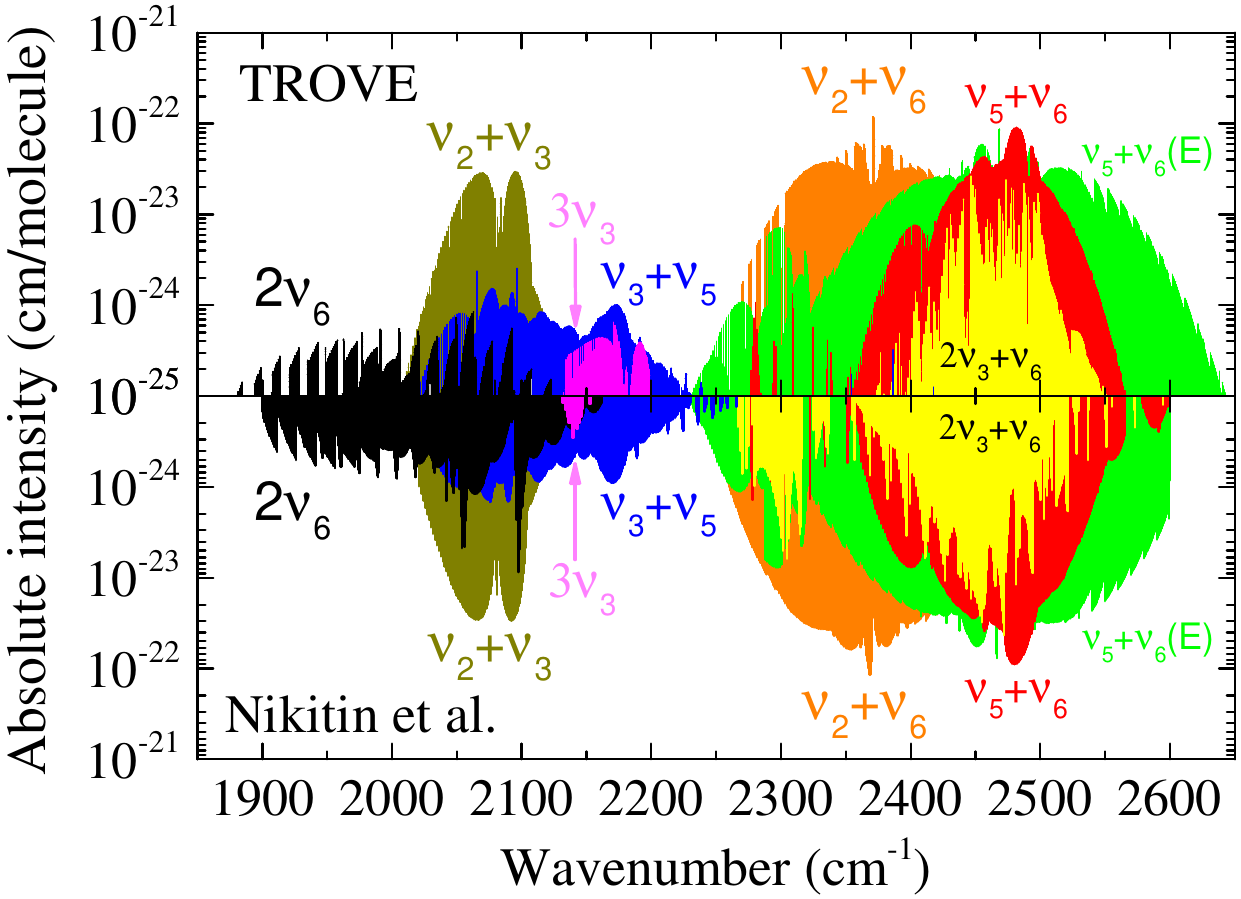}
\caption{\label{fig:nikitin}Absolute line intensities of CH$_3{}^{35}$Cl in the range $1900${--}$2600\,$cm$^{-1}$ compared with measurements from \citet{16NiDmGo.CH3Cl}. Computed TROVE transitions are up to $J=85$ whilst the results from \citet{16NiDmGo.CH3Cl} are up to $J=47$. Note that a logarithmic scale has been used for the y-axis.}
\end{figure}

 An improved spectroscopic line list in the range $1900${--}$2600\,$cm$^{-1}$ was recently published~\citep{16NiDmGo.CH3Cl} and considered transitions up to $J=47$ with absolute line intensities possessing an estimated uncertainty of $20\%$ or less. In Fig.~\ref{fig:nikitin} a comparison of this region, which is composed of several weak bands, is shown for CH$_3{}^{35}$Cl. The DMS appears reasonable for much weaker intensities and the overall band structure in this region is well reproduced. There are some irregularities between TROVE and \citet{16NiDmGo.CH3Cl}; we expect these are caused by the low-level nature of our calculations and also the assignment procedure in TROVE. In future work we intend to carry out a more comprehensive analysis of this region. Note that the computed TROVE line list has not been truncated at $J=47$ for this comparison.

\begin{figure}
\includegraphics[width=\textwidth,angle=0]{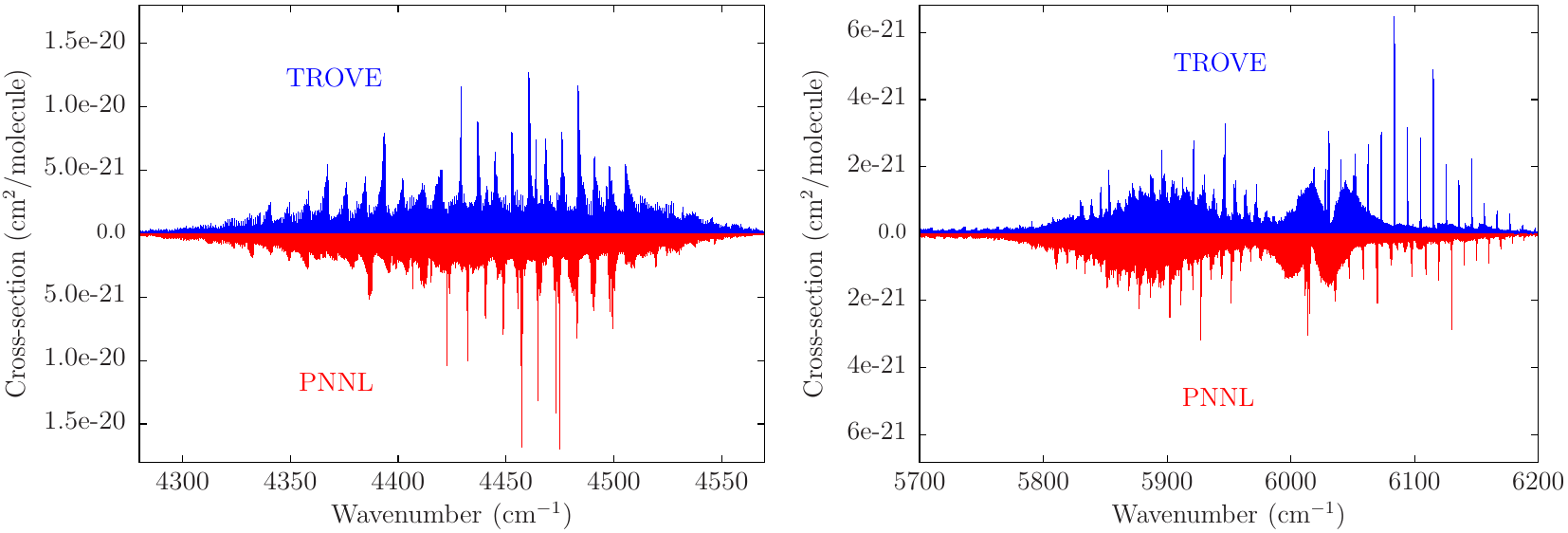}
\caption{\label{fig:pnnl}Overview of simulated rotation-vibration spectrum in the $4300${--}$4550\,$cm$^{-1}$ (left) and $5700${--}$6200\,$cm$^{-1}$ (right)  regions compared with the PNNL spectral library~\citep{PNNL}. Computed intensities have been scaled to natural abundance.}
\end{figure}

\begin{figure}
\includegraphics[width=\textwidth,angle=0]{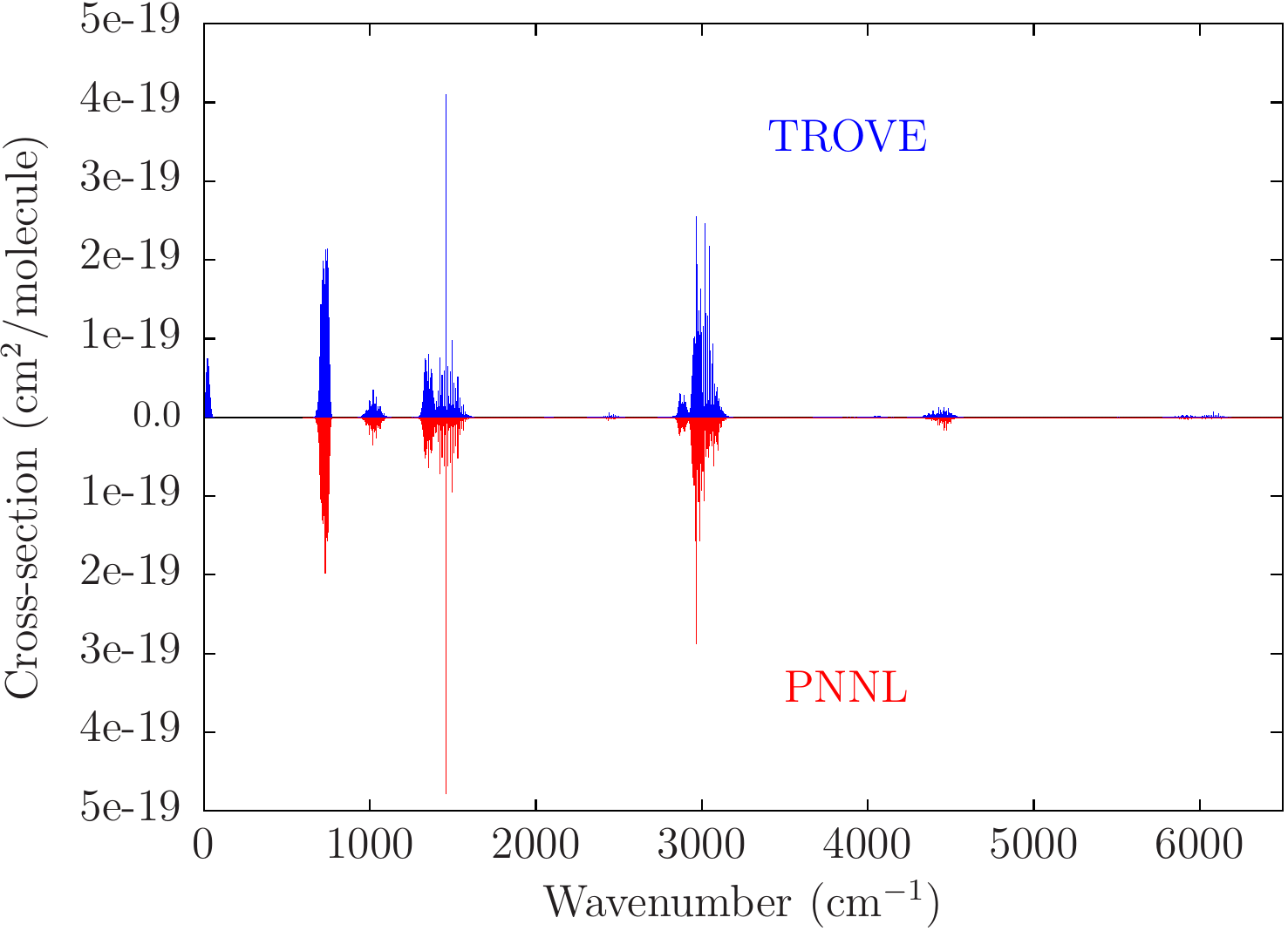}
\caption{\label{fig:all_pnnl}Overview of methyl chloride rotation-vibration spectrum up to $J=85$ compared with the PNNL spectral library~\citep{PNNL}. Computed intensities have been scaled to natural abundance.}
\end{figure}

 Whilst spectral features above $3200\,$cm$^{-1}$ are not as prominent there are noticeable bands between $4300${--}$4550\,$cm$^{-1}$ and $5700${--}$6200\,$cm$^{-1}$ as shown in Fig.~\ref{fig:pnnl}. Here we have compared against the PNNL spectral library~\citep{PNNL} (overview of entire spectrum presented in Fig.~\ref{fig:all_pnnl}). Cross sections have been generated at a resolution of $0.06{\,}$cm$^{-1}$ and fitted using a Gaussian profile with a half width at half maximum of $0.112{\,}$cm$^{-1}$. This line shape provides a straightforward and reasonable comparison~\citep{15AlYaTe.H2CO}, however, we expect a Voigt profile dependent on instrumental factors would be more suitable. 

 Looking at Fig.~\ref{fig:pnnl} it is clear that our calculations are becoming worse at higher energies and producing spurious intensities. This is to be expected given the size of the vibrational basis set and thresholds imposed in our variational calculations. For the complete spectrum in Fig.~\ref{fig:all_pnnl} the agreement with PNNL is encouraging but does indicate the need for improved computations and line shape modelling.

\section{Conclusions}
\label{sec:conc}

 A new nine-dimensional DMS for methyl chloride has been computed using high-level \textit{ab initio} theory and fitted with a symmetry-adapted analytic representation. The DMS has been utilized to simulate the rotation-vibration spectrum of methyl chloride up to $6300\,$cm$^{-1}$. Overall, band shape and structure are well reproduced and there is good agreement with a range of experimental sources. Notably, at least for fundamental bands that have been measured with high accuracy, computed absolute line intensities agree well with the available experimental data.

 Considering the quality of the DMS and intensity simulations, further improvements could come from using a larger (augmented) basis set for the electronic structure calculations. However, this would be very computationally demanding and the change in predicted intensities may not necessarily reflect the computational effort. Empirical refinement of the PES~\citep{YuBaTe11.NH3} is expected to produce more reliable intensities~\citep{OvThYu08a.PH3} (as a result of better rovibrational energy levels and associated wavefunctions) and should thus be done when attempting to produce a line list suitable for high-resolution spectroscopy.

 Most beneficial though will be improvements in the variational nuclear motion calculations. This will require increasing both the size of the vibrational basis set and the frequency range considered. Measurements of absolute line intensities of the order $10^{-26}${--}$10^{-27}\,$cm/molecule have been reported in the $11\,590${--}$11\,760\,$cm$^{-1}$ spectral region~\citep{16LuGoxx.CH3Cl}. Presently we are unable to accurately model such high frequencies; this is a major challenge for variational calculations on small polyatomic molecules. We expect these issues to be addressed during the construction of a comprehensive, hot line list for inclusion in the ExoMol database~\citep{ExoMol2012,ExoMol2016}.

\section*{Acknowledgements}

The authors are grateful to Agnes Perrin and David Jacquemart for discussing results concerning their intensity measurements. This work was supported by ERC Advanced Investigator Project 267219, and FP7-MC-IEF project 629237.

\bibliographystyle{elsarticle-num-names}

\end{document}